\documentclass[%
 reprint,
 amsmath,amssymb,
 aps,
]{revtex4-2}

\usepackage{graphicx}
\usepackage{dcolumn}
\usepackage{bm}

\begin{document}

\preprint{APS/123-QED}

\title{
Fluidity in Domain Walls in Dilute $^3$He-$^4$He Films on Graphite:\\
Possible 1D Fermi Fluid and Dirac Fermions in Helium Film
}

\author{Masashi Morishita}
\email{morishita.masashi.ga@u.tsukuba.ac.jp}
\affiliation{%
Pure and Applied Sciences, University of Tsukuba, Tsukuba, Ibaraki 305-8571, Japan
}%




\date{\today}

\begin{abstract}
The heat capacity of a small amount of $^3$He atoms dissolved in submonolayer $^4$He film has been measured. 
The measured heat capacity is finite and suggests that $^3$He atoms are mobile at an areal density regime higher than that of the $\sqrt{3}\times\sqrt{3}$ phase, where $^4$He films are believed to be solid.
At higher areal densities, the measured heat capacity is proportional to $T^2$ and depends on the amounts of $^3$He atoms. 
These behaviors are anomalous to that of a two-dimensional Fermi fluid, and cannot be explained by uniform melting of $^4$He films. 
One possible explanation for these anomalous behaviors is that helium atoms exhibit fluidity only inside the domain walls of the adsorption structure, and the dissolved $^3$He atoms gather into them and behave as a one-dimensional Fermi fluid or as Dirac fermions, depending on the structure of the domain walls. 
The behaviors of the measured heat capacity strongly suggest this possibility.


\end{abstract}

\maketitle



The quantum properties of low dimensional matter have attracted much attentions in condensed matter physics. 
Graphene is one of the most fascinating and peculiar examples that can be treated as two-dimensional (2D) \cite{[A. H. Castro ]CastroNeto}, because it exhibits novel and unique features, and studies on its properties and applications have evolved explosively in the last decade. 
A helium film adsorbed onto a graphite surface provides an almost ideal 2D system, and exhibits a well-defined layer-by-layer structure \cite{Zimmerli}.
Each layer is independent from each other and exhibits high flatness and uniformity.
The $^3$He atom has a nuclear spin of 1/2, and $^3$He solid film provides a 2D quantum spin system and has been investigated vigorously \cite{Godfrin-Lauter,Godfrin-Rapp,Fukuyama2008}.
With increase in areal density, their magnetism exhibit rather complicated change \cite{GreywallPRB41, Godfrin1994, Ikegami1998}, which has been discussed with the evolution of the adsorption structure \cite{Morishita-Takagi2001, Morishita-Takagi2003}.
On the other hand, information on the properties and adsorption structures of helium-4 ($^4$He) films is limited due to the lack of an appropriate method for their observation. 

In this letter, I report the results of heat capacity measurements of a small amount of $^3$He atoms dissolved into submonolayer $^4$He films on graphite.
Results strongly suggest that $^3$He atoms are mobile at an areal density regime higher than that of the $\sqrt{3} \times \sqrt{3}$ phase, where $^4$He films have been believed to be solid.
At higher areal densities the measured heat capacity is in proportion to the square of temperature.
This anomalous temperature variation cannot be explained if $^3$He atoms move around the entire surface of the graphite.
At these areal densities, $^4$He films are expected to have domain wall superstructures \cite{ Halpin-Healy, Greywall1993, Mohandas, Corboz2008}.
A possible explanation for these anomalous observations is that $^4$He atoms in domain walls exhibit fluidity, and that $^3$He atoms gather into and move around only inside domain walls.
Fluidity inside domain walls provides regular confined geometry with the width of atomic size for $^3$He atoms. 
$^3$He atoms can be expected to behave as a one-dimensional Fermi fluid or as Dirac fermions, depending on the structure of domain walls (striped or honeycomb).

The heat capacity measurement of dilute $^3$He-$^4$He mixture films can be utilized to clarify the nature of $^4$He films, and this approach was first adopted by Ziouzia {\it et al.} \cite{Ziouzia2003}. 
The heat capacity of $^4$He is very small \cite{Greywall1991,Greywall1993}, and gives very little information about the nature of $^4$He films.  
A small amount of $^3$He atoms dissolve only into the top layer of $^4$He thin film. 
When the top layer of $^4$He film is a fluid, the $^3$He atoms behave as a Fermi fluid and exhibit finite heat capacity, giving information on the $^4$He film. 
On the other hand, when the $^4$He film is solid, the dissolved $^3$He are almost localized and exhibits almost no heat capacity contribution. 

The heat capacity is measured by the usual adiabatic heat-pulse method. 
The graphite substrate used in this work is Grafoil. 
The total surface area of the substrate is approximately 390 m$^2$. 
To ensure uniformity of $^3$He-$^4$He film, the following procedures are adopted in sample preparation. 
At first, a sufficient amount of the sample $^4$He is introduced into the sample cell to cover the heterogeneous surface of the Grafoil substrate.
After the $^4$He film is annealed by raising the temperature once, a designated amount of $^3$He gas is introduced, and the sample film is annealed again.
Typically, in a series of measurements, the amount of $^3$He is fixed at some value which corresponds to the areal densities ($\rho_3$) of 0.1 nm$^{-2}$ or 0.2 nm$^{-2}$, while the amount of $^4$He is gradually increased. 
Annealing is performed after the introduction of each sample over 6-8 h at a high temperature with the sample vapor pressure at around 500 Pa. 
After the annealing, the temperature is slowly decreased over 8-10 h until the vapor pressure becomes much less than 1 Pa. 
The vapor pressure is measured by an in-situ pressure gauge.
Other experimental details are similar to those of our previous works \cite{Morishita2013, Morishita2002}.

\begin{figure}[b]
\includegraphics [viewport= .0cm .0cm 13.5cm 23.5cm, clip, width=.85\linewidth, keepaspectratio]{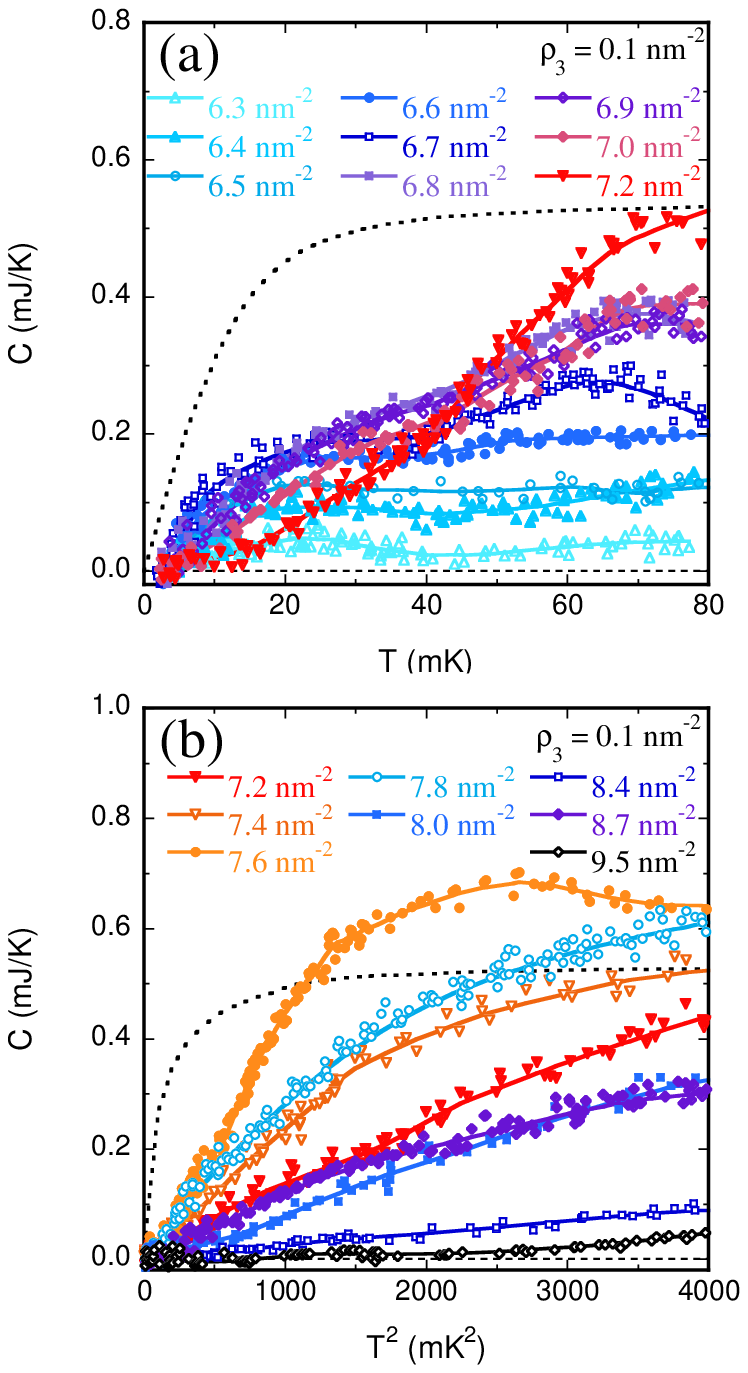}
\caption{
(color online). 
Measured heat capacity of dilute $^3$He-$^4$He mixture films are plotted at select areal densities (a) below 7.2 nm$^{-2}$ as functions of $T$, and (b) above 7.2 nm$^{-2}$ as functions of $T^2$ (b).
Numbers in the figure indicate the total areal density of $^3$He and $^4$He ($\rho_{total}$), and here the areal density of $^3$He is 0.1 nm$^{-2}$. 
The dotted lines indicate the expected heat capacity of ideal 2D Fermi gas of 0.1 nm$^{-2}$, and the broken lines indicate the origin of the vertical axes. 
The solid lines are guides for the eye. 
}
\label{fig:rawdata}
\end{figure}


The measured heat capacities ($C$) of submonolayer dilute $^3$He-$^4$He mixture films at select areal densities are shown in Fig. \ref{fig:rawdata} as functions of temperature ($T$) or of the square of temperture ($T^2$).
As reported elsewhere \cite{Morishita2016}, with increasing areal density from the fluid phase and approaching the areal density of 6.3 nm$^{-2}$, the measured heat capacity approaches zero.
This value of areal density corresponds to that of the $\sqrt{3} \times \sqrt{3}$ phase, and this behavior can be attributed to the solidification of the $^3$He-$^4$He film into the $\sqrt{3} \times \sqrt{3}$ phase. 
However, with further increase of the areal density, the measured heat capacity increases and becomes finite, as shown in Fig. \ref{fig:rawdata}.
This behavior suggests that the $^3$He atoms are mobile, although at this areal density regime, the $^4$He film is believed to be solid. 
Furthermore, at the higher areal densities than 7.2 nm$^{-2}$ the measured heat capacity is proportional to $T^2$ (as shown in Fig. \ref{fig:rawdata}(b)) and the magnitude of the measured heat capacity is almost proportional to the amount of $^3$He atoms (as shown below in Fig. \ref{fig:gamma2}.).
The heat capacity of a Fermi fluid is proportional to $T$ at low temperatures, and its slope is independent of the number density of particles. 
Therefore, $^3$He atoms dissolved in submonolayer $^4$He films at these areal densities cannot be considered a Fermi fluid, and uniform melting of $^4$He films cannot explain the observations.

There are some candidates for the possible origins of the observed anomalous heat capacity. 
The $T^2$ variation reminded us of 2D phonon contribution. 
However, the heat capacities of pure $^4$He films, whose origins can be attributed to phonons, are far smaller than the measured ones here \cite{Greywall1991, Greywall1993, Morishita2017}. 
In other words, the magnitude of the observed heat capacity can be explained only by the non-realistic Debye temperature of the order of 1 mK. 
$^3$He nuclear spin contribution can also be excluded, 
because the interactions between $^3$He nuclear spins should be extremely weak in the context of this experiment, 
and furthermore, entropy changes calculated from measured heat capacities are much larger than the expected change, $N_3k_B\ln2$, where $N_3$ is the number of $^3$He atoms.
A film consisting of a $^3$He-$^4$He mixture can exhibit phase separation into $^3$He-rich and $^4$He-rich phases, and the mixing of these phases with increasing temperature is also a candidate for the origin of the observed heat capacity.
However, heat capacity contribution from the mixing should be independent of the amount of $^3$He. 
The $^3$He amount dependence of the observed heat capacity can exclude this possibility.

Helium films are thought to solidify with the important contribution of hardcore repulsion between helium atoms \cite{Casey2003,Hirashima,Matsumoto2005}. 
At higher areal densities, a plausible structure is the domain wall (DW) superstructure. 
DWs exhibit two different structures, namely the striped and honeycomb DW structures. The DW structures have been discussed and observed in many adsorbed systems on graphite. 
For $^3$He monolayer film, areal density evolution of the DW structures has been discussed according to Monte Carlo calculations \cite{Morishita-Takagi2001,Morishita-Takagi2003}. 
Also for $^4$He monolayer film on graphite, DW structures have been theoretically predicted \cite{Halpin-Healy, Corboz2008} and proposed according to experimental observations \cite{Greywall1993}. 
In the DWs, the role of the corrugation on solidification is less important, and $^4$He could exhibit fluidity. 
The situation is somewhat similar to the possible fluidity inside dislocations and grain boundaries in hcp $^4$He concerning its observed ``supersolid''-like behavior \cite{Chan2013}. 
If DWs exhibit fluidity, $^3$He atoms should crowd onto the DWs to reduce their zero-point energies, and move about in the DWs.
Therefore, confined geometries with the width of atomic size are provided for $^3$He atoms.  
Although the structures of dislocations and grain boundaries in hcp $^4$He are irregular, the DWs are arranged regularly, and the behaviors of $^3$He atoms dissolved in them can be expected to reflect the regular structure. 

In the case of striped DW structures (which appear in a lower areal density regime), $^3$He atoms should travel in one-dimensional (1D) space and behave as a 1D Fermi fluid.
Hence, $^3$He atoms dissolved in striped DWs in $^4$He film are a possible candidate for a Tomonaga-Luttinger liquid, although evidence for this is yet to be obtained in heat capacity measurements.

In the case of honeycomb DW structures (which appear in a higher areal density regime), $^3$He atoms should travel in honeycomb lattices. 
Their degree of freedom is similar to that of electrons in graphene. 
In graphene, electrons behave as massless Dirac fermions and their dispersion near the Dirac points is linear \cite{[A. H. Castro ]CastroNeto}. 
Similar behaviors have been observed in an ultracold gas of potassium atoms in honeycomb lattices \cite{Tarruell}, and in carbon monoxide molecules in a hexagonal pattern \cite{Gomes}. 
$^3$He atoms in the honeycomb DWs of $^4$He films are similarly expected to have linear dispersion. 
In this case, their heat capacity is expected to be proportional to $T^2$, and observed anomalous $T^2$ variation at high areal densities can be explained. 
An almost $T^2$ dependence has been reported in the heat capacity measurement of a multilayered organic material, in which massless Dirac fermions are expected \cite{Konoike}.

\begin{figure}[b]
\includegraphics[viewport= 1.8cm 8.8cm 18.5cm 20.2cm, clip, width=0.85\linewidth,keepaspectratio]{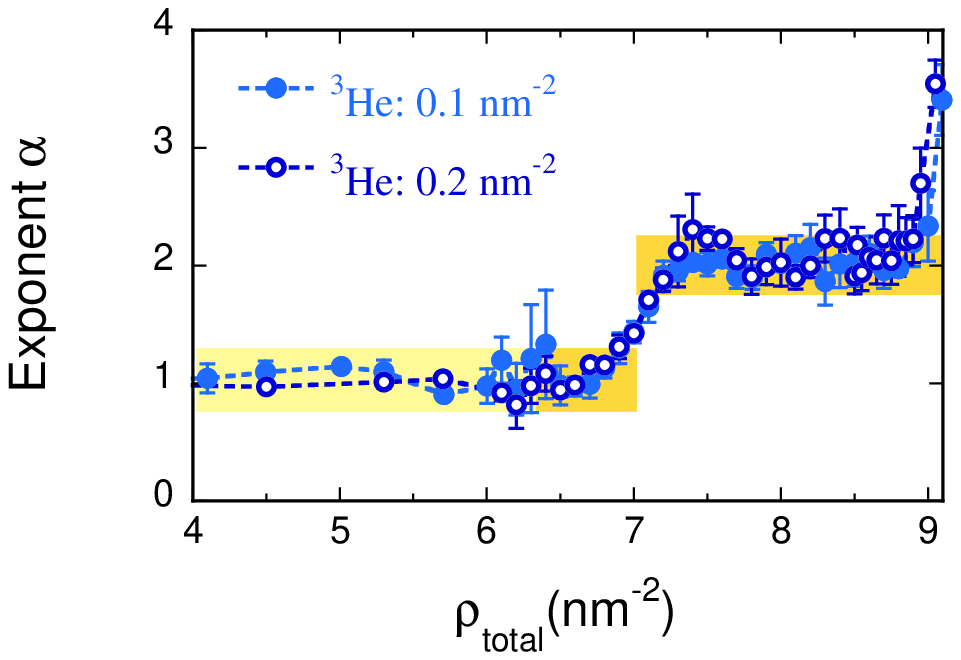}
\caption{
(color online). 
Exponent of measured heat capacity of dilute $^3$He-$^4$He mixture films at a low temperature regime. 
The colored rectangles are guides for the eye. 
Sudden change at around 7.0 nm$^{-2}$ strongly suggest the structural phase transition between striped and honeycomb domain wall structures. 
}
\label{fig:beki}
\end{figure}


The linear dispersion relation of graphene is usually explained using the tight-binding approach \cite{CastroNeto, Wallace}, which is not adequate for the $^3$He atoms dissolved into $^4$He fluid. However, it has been revealed that the tight-binding nature is not necessary for the appearance of the linear dispersion relation. Lomer discussed the band structure of graphene using a group-theoretical treatment \cite{Lomer}. Park and Louie showed that the Dirac fermion nature can be generated within an independent particle picture \cite{Park}. Geng et al. found that the linear dispersion relation appears in discretized tight-binding models, although the Fermi level does not coincide with the Dirac point \cite{Geng, Geng_thesis}.


The exponent ($\alpha$) of the measured heat capacity, which is obtained by fitting the measured values with $C\propto T^\alpha$ in the low temperature regime, where the second derivative of the smoothed values is not negative, is plotted in Fig. \ref{fig:beki}.
Results shown in Fig. \ref{fig:beki} include those of films in a low areal density regime where films are fluid. 
The rather sudden change from $T$-linear to $T^2$ behavior at around 7.0 nm$^{-2}$ can be attributed to the structural phase transition between the striped  and honeycomb DW structures. 
In the case of submonolayer pure $^3$He film, the structural phase transition between striped  and honeycomb DW structures is predicted to occur around 6.8 nm$^{-2}$ \cite{Morishita-Takagi2003}. 
This value is similar to that of the areal density where the exponent of the measured heat capacity suddenly changes, although the masses and the quantum statistics are different between $^3$He and $^4$He. 


Next, let us pay attention to the behaviors in a high temperature regime. 
The heat capacity of a 1D Fermi fluid approaches $N_3k_B/2$ at the high temperature limit. 
In Fig. \ref{fig:rawdata}, the dotted lines indicate the expected behavior for a 2D Fermi gas which saturates to $N_3k_B$. 
In Fig. \ref{fig:rawdata}(a), the measured heat capacities tend to saturate to $N_3k_B/2$ at high temperatures.  
The observed smaller values can be attributed to the finite solubility of $^3$He in domain walls. 
That is, some fraction of $^3$He atoms dissolves in $\sqrt{3} \times \sqrt{3}$ domains and is localized. 
On the other hand, at areal densities between 6.7 and 6.9 nm$^{-2}$, the measured heat capacities tend to saturate once to $N_3k_B/2$ at around 40 mK, but increase further at higher temperatures. 
The DW structure may change from striped to honeycomb with increasing temperature near the critical areal density.

With increasing temperature, the heat capacity of a 2D Fermi gas with linear dispersion is expected to overshoot once and then decrease and saturate to $2Nk_B$, which is twice the expected value for an ordinary 2D Fermi gas. 
The measured heat capacities appear to approach $N_3k_B$, and not $2N_3k_B$, with a rather large distribution. 
The thermal de Broglie length of $^3$He atoms at 10 mK is about 10 nm, which is similar to the platelet size of Grafoil \cite{Birgeneau} ; 
at 100 mK it is several nm, which is similar to the lattice constants of the honeycomb domain wall structures. 
Therefore, in a sufficiently low temperature regime, $^3$He atoms can be affected by the honeycomb structures. 
However, at higher temperatures, $^3$He atoms should behave as ordinary 2D fermions, and their heat capacity should approach $N_3k_B$. 
The excess observed at around 7.6 nm$^{-2}$ can be explained by that the heat capacity exceeds $N_3k_B$ before the $^3$He atoms loose the nature of Dirac fermions, or the linear dispersion, with increasing temperature. 
Conversely, the observation of the excess supports the peculiarity of this system. 
Indeed, at these areal densities, the measured heat capacities tend to decrease and seemingly approach $N_3k_B$ at high temperatures.

\begin{figure}[t]
\includegraphics[viewport= 2cm 10.cm 19cm 19cm, clip, width=.85\linewidth,keepaspectratio]{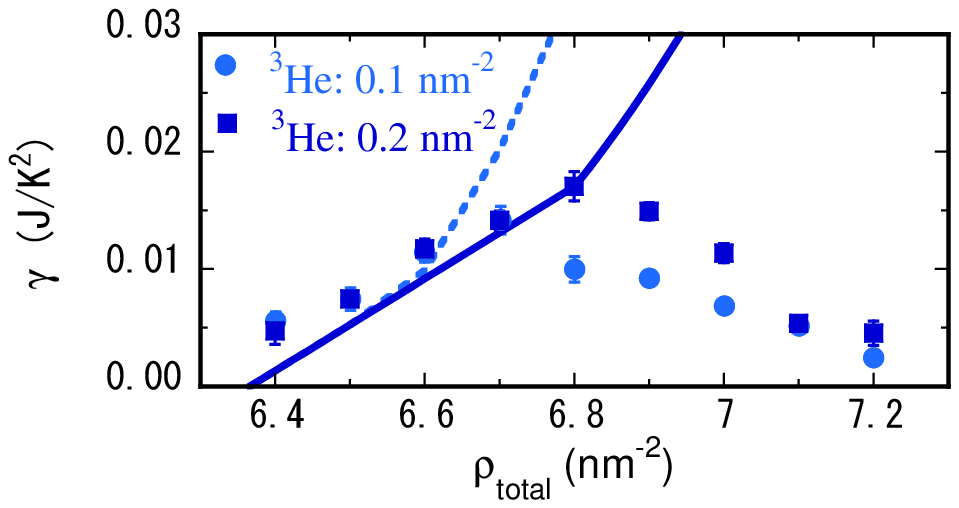}
\caption{
(color online). 
Areal density variation of the slope of measured heat capacity. 
The broken line and the solid line are expected behavior for the striped domain wall structure with $\rho_3 =$ 0.1 and $\rho_3 =$ 0.2 nm$^{-2}$, respectively, with assumptions described in the text. 
}
\label{fig:gamma1}
\end{figure}

The slope of the heat capacity of 1D Fermi gas at low temperatures is 
$\gamma=g^2k_B^2mL^2/3\hbar^2N$,
where $g$ is the number of degrees of spin freedom, $m$ is the mass, and $L$ is the length. 
Unfortunately, the number of $^3$He atoms in striped DWs depends on the (total) areal density due to finite solubility as mentioned above. 
However, the solubility can be thought of as almost proportional to the length of the DWs, and the total length of the striped DWs increases linearly with the (total) areal density. 
The expected changes in $\gamma$, assuming the solubility of $^3$He reaches a value corresponding to 0.2 nm$^{-2}$ at the total areal density of 6.8 nm$^{-2}$, are shown in Fig. \ref{fig:gamma1} for cases with $\rho_3 = $ 0.1 and 0.2 nm$^{-2}$. 
The slopes of the measured heat capacity, which are obtained by linear fitting of the measured heat capacity at low temperatures (typically below 20 mK), are also shown in Fig. \ref{fig:gamma1}. 
The agreement with the expected areal density variation is good at low areal densities. 
The decreases in the slopes of the measured heat capacity at high areal densities can be attributed to the coexistence of honeycomb DWs. 

\begin{figure}[t]
\includegraphics[viewport= 0cm 7cm 20cm 20cm, clip, width=0.85\linewidth,keepaspectratio]{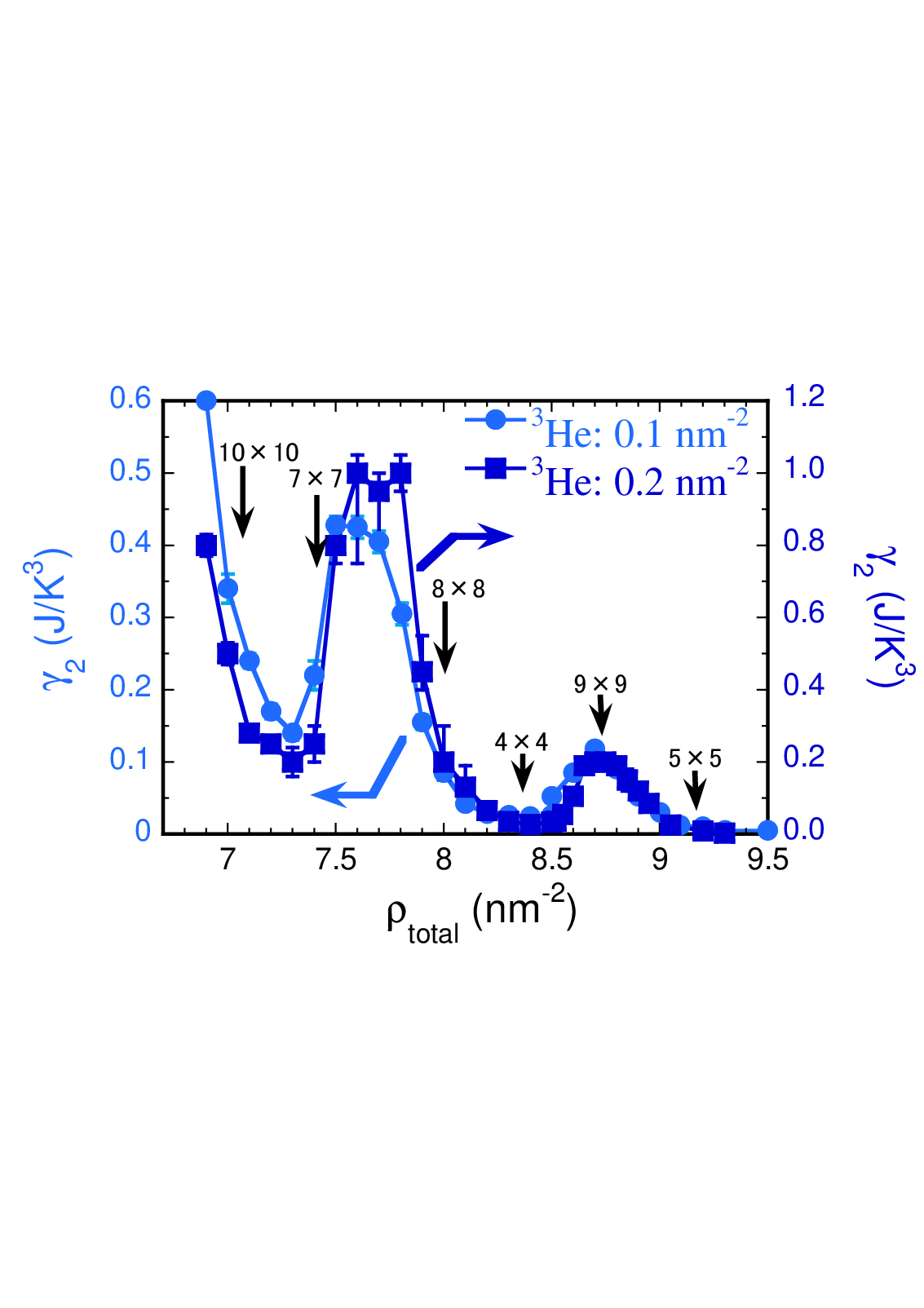}
\caption{
(color online). 
Areal density variation of the coefficient of the $T^2$ term of the measured heat capacities.
The scale of vertical axis for $\rho_3$ = 0.1 nm$^{-2}$ is shown on the left, and for $\rho_3$ = 0.2 nm$^{-2}$ on the right.
These scales differ from each other according to the amount of $^3$He.
The arrows indicate the areal densities where the honeycomb domain wall structure with displayed periodicity has a regular hexagonal structure. 
}
\label{fig:gamma2}
\end{figure}


The coefficients of the $T^2$ term, $\gamma_2$ is obtained by fitting the measured values with $C=\gamma_2T^2$ at a low temperature regime (typically below 30 mK).
As shown in Fig. \ref{fig:gamma2}, the $T^2$ term disappears at approximately around 9.1 nm$^{-2}$. 
Further, as depicted in Fig. \ref{fig:beki}, the $T^2$ term disappears at approximately 8.8 nm$^{-2}$.
These observations indicate that honeycomb DW structures survive up to considerably higher areal densities than the expected values––7.9 nm$^{-2}$ from heat capacity measurements \cite{Greywall1991} or 8.4 nm$^{-2}$ from theoretical simulations \cite{Corboz2008}. 

\begin{figure}[t]
\includegraphics[viewport= 1.8cm 8.5cm 18.5cm 21.5cm, clip, width=0.85\linewidth,keepaspectratio]{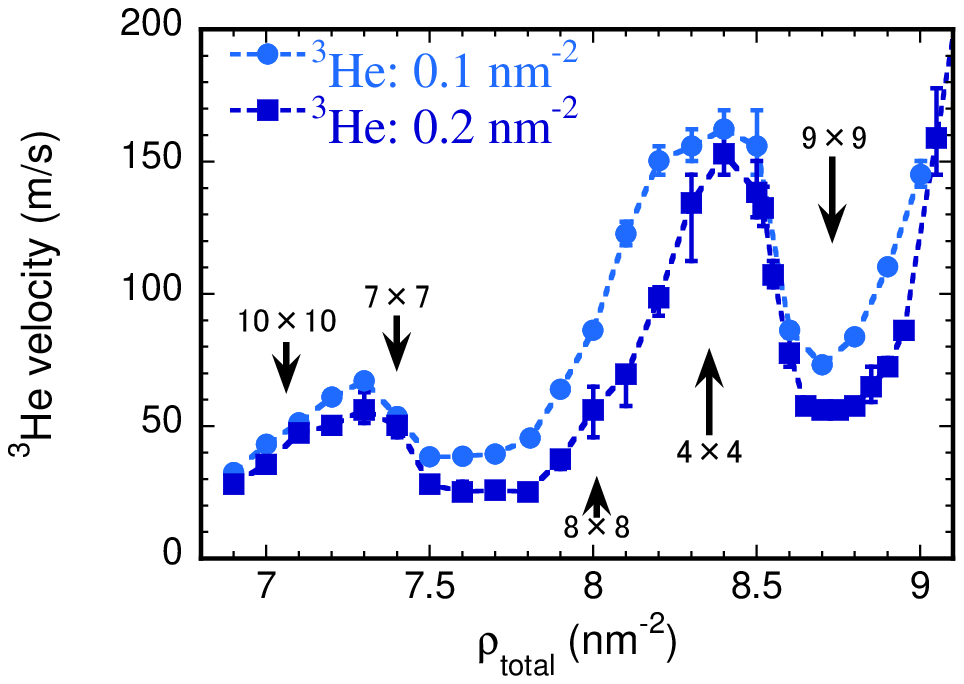}
\caption{
(color online). 
Areal density variation of the velocity of $^3$He atoms estimated from the measured heat capacity, assuming $^3$He atoms behave as Dirac fermions.
The increase observed at areal densities above 8.6 nm$^{-2}$ may be incorrect due to the coexistence of domain wall structure and incommensurate solid phase.
}
\label{fig:speed}
\end{figure}


If $^3$He atoms behave as Dirac fermions, their speeds are the same, because they exhibit linear dispersion. 
The speed of $^3$He atoms, $v_3$, can be estimated from $\gamma_2$ with the following formula, 
$v_3= \left(9g\zeta(3)k_B^3A / 2\pi\hbar\gamma_2 \right)^{1/2}$, 
assuming that interactions between $^3$He atoms are weak, and where $g=4$ is the number of degrees of degeneracy, $\zeta(3)$ is the Riemann zeta function, and $A$ is the surface area \cite{Vafek}.
The estimated velocities are shown in Fig. \ref{fig:speed} as functions of total areal density.
The estimated velocity  has maxima at around 8.4 nm$^{-2}$, and it is much higher than the Fermi velocity in $^3$He films behaving as 2D Fermi fluid.
At 8.4 nm$^{-2}$, the honeycomb DW structure is expected to have a regular structure with the periodicity of $4\times4$ against the periodicity of the hollow sites of graphite.
This structure has the smallest periodic length within honeycomb DW structures. 
Therefore, the honeycomb structure can be defined very well here, although the platelet size of Grafoil is small. 
The maximum value of $v_3$ at this areal density can be attributed to this reason.

The obtained magnitude of $v_3$ appears to saturate at approximately 160 m/s. 
This behavior suggests the existence of some critical velocity. 
Although one possible origin is the critical velocity of the 2D superfluid $^4$He, measurements with smaller amounts of $^3$He are desirable.


In summary, the heat capacity of a small amount of $^3$He atoms dissolved in submonolayer $^4$He film on graphite was measured. 
The observed behaviors suggest the nature of $^3$He atoms to be that of 1D fermions in the low areal density regime, and that of Dirac fermions in the higher areal density regime. 
These results 
strongly suggest that the films exhibit fluidity in the domain walls. 
The origin of the fluidity and natures of $^3$He and also $^4$He atoms in domain walls must be understood further with successive research.

\begin{acknowledgments}
The author acknowledges stimulating discussions with Takeo Takagi, Tomoki Minoguchi, Masaki Oshikawa, Chenhua Geng, Yasuhiro Hatsugai, and Yoichi Ootuka. 
This research was supported by the Cryogenics Division, Research Facility Center for Science and Technology, University of Tsukuba.
\end{acknowledgments}

\bibliography{Dirac3}   

\end{document}